\documentclass[aps,twocolumn]{revtex4}

\usepackage{graphpap}

\usepackage{graphicx}
\usepackage{color}

\begin{document}

\title{Transmission through a quantum dot molecule embedded in an Aharonov-Bohm interferometer}

\author{Daniel A. Lovey}

\author{Sergio S. Gomez}

\affiliation{Instituto de Modelado e Innovaci\'on Tecnol\'ogica, CONICET, and\\
Facultad de Ciencias Exactas y Naturales y Agrimensura, Universidad Nacional del Nordeste, \\
Avenida  Libertad 5500 (3400) Corrientes, Argentina.}

\author{Rodolfo H. Romero}
\email{rhromero@exa.unne.edu.ar}
\affiliation{Instituto de Modelado e Innovaci\'on Tecnol\'ogica, CONICET, and\\
Facultad de Ciencias Exactas y Naturales y Agrimensura, Universidad Nacional del Nordeste, \\
Avenida  Libertad 5500 (3400) Corrientes, Argentina.}

\date{ \today}

\begin{abstract}
We study theoretically the transmission through a quantum dot molecule embedded in the arms of an Aharonov-Bohm four quantum dot ring threaded by a magnetic flux. The tunable molecular coupling provides a transmission pathway between the interferometer arms in addition to those along the arms. 
From a decomposition of the transmission in terms of contributions from paths, we show that antiresonances in the transmission arise from the interference of the self-energy along different paths and that application of a magnetic flux can produce the suppression of such antiresonances. 
The occurrence of a period of twice the quantum of flux arises to the opening of transmission pathway through the dot molecule. 
Two different connections of the device to the leads are considered and their spectra of conductance are compared as a function of the tunable parameters of the model.
\end{abstract}

\maketitle

\section{Introduction}
The reduction of size in the electronic devices to nanometer scale has highlighted the importance of the effects of quantum coherence and interference. The control of such effects is important to provide both a better understanding of the quantum realm as well as new functionalities to the circuits \cite{Imry02, DiVentra08}. 
Quantum interference allows to enhance or to cancel, total or partially,  the response of the system beyond the simple classical additive behaviour.  Such an effect can pose a problem to be avoided but also could provide new capabilities to the device with respect to its classical counterpart \cite{Aronov87, Hod08}. There is currently an increasing interest in being able to tune the parameters of mesoscopic systems, what would enable to manipulate their quantum behaviour for the advantageous design and applications of future electronic devices \cite{Barnham01}. 

The characteristic quantum phenomenon of the change of phase of the wave function along two paths enclosing a magnetic flux, i.e. the Aharonov-Bohm (AB) effect \cite{Aharonov59}, has been envisioned for the feasible exploitation of the quantum phase in electronic devices \cite{Hod08, Hackenbroich01, Hod04, Hod06}. Phase coherent effects in AB rings have been treated theoretically in the literature \cite{Guevara03, Guevara06, Gomez04, Hedin09, An10, Guevara10}.
Closely related to the concept of coherence is the quantum interference between a discrete state with a continuum of states. This phenomenon was firstly studied by Fano in the spectrum of photoionization of atoms and termed Fano effect after him \cite{Fano61}, but was found ubiquitous in a wide range of physical phenomena. Interestingly, it has been observed in properly tailored nanoscale systems \cite{Kobayashi02, Fuhrer06, Ihn07, Miroshnichenko10}. The first tunable experiment showing the characteristic asymmetric Fano profile in the electron transmission was reported in an AB ring with a quantum dot embedded in one of its arms \cite{Kobayashi02}. Various parameters, such as the gate voltage and the magnetic flux through the ring among others, allowed to tune the peak and dip of the profile. More recently, a quantum dot molecule, i.e. two coupled quantum dots, has been embedded between the arms of an AB interferometer, with even a larger number of tunable parameters \cite{Ihn07}; the transmission thus exhibits a large variety of behaviours as a function of them. In \cite{Ihn07} it is stated that the simplest fully coherent single-mode picture is of limited use for the interpretation of their experiments. In such a picture, the transmission is assumed to arise from a direct reflection and another one after travelling once around the AB ring. It is the purpose of this paper to improve the theoretical description of the experimental results.
\\ \indent
Here we present theoretical calculations of a model inspired in the system of Ref. \cite{Ihn07} within a phase coherent non-interacting formalism. On one hand, we assessed the sensitivity of the conductance upon variation of the various model parameters. On the other hand, we show that the self-energy matrix elements, obtained from applying a partitioning technique to the Hamiltonian, can give insight on the quantum interference and its dependence on the magnetic flux through the ring. This interpretation in terms of contributions through pathways, allows one to predict the onset or cancellation of the antiresonances as a function of the parameters of the model.
\\ \indent
This paper is organized as follows: Section \ref{model} describes the model of the device and its connections to the leads. 
Section \ref{section partitioning} introduce the spacial contributions to the conductance by means of a partitioning technique and discusses the conditions for the onset of peaks and cancellation of conductance in the transmission in terms of the Green functions of the isolated device. In Section \ref{section results} we present the results obtained varying the various model parameters; finally, in Section \ref{section conclusions} we summarize our conclusions.
\section{\label{model} Model}
\begin{figure}[h]
\includegraphics[scale=0.3]{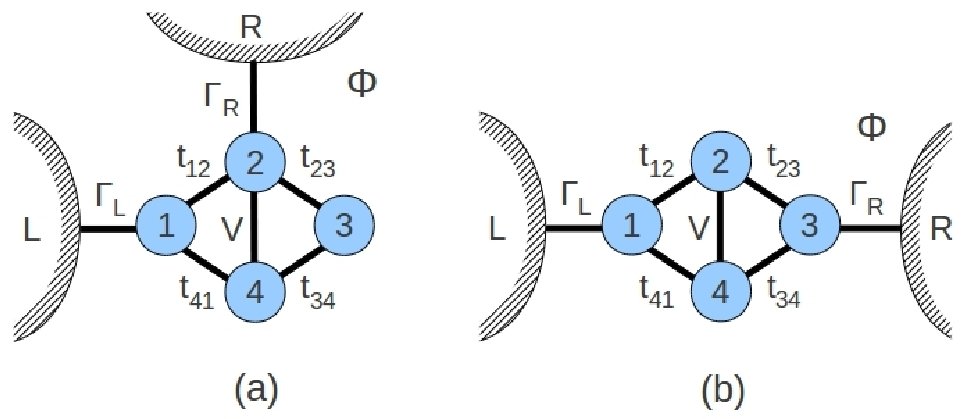} 
\caption{\label{system} Scheme of the device: (a) in the (1,2)-connection and (b) (1,3)-connection to the leads.}
\end{figure}

We consider four quantum dots forming a ring and coupled to two leads L and R. Sites 2 and 4 of the ring are connected to each other forming an artificial molecule, as shown in Fig. \ref{system}.
 We consider only one energy level in each dot and both the intradot and interdot electron-electron interactions are neglected. The system shall be described by a Hamiltonian
\begin{equation}
H = H_r + H_l + H_t,
\end{equation}
where $H_r$ is the Hamiltonian of the isolated bicyclic ring, 
\begin{eqnarray}
H_r = \sum_{i=1}^4 \varepsilon_i d_i^\dag d_i &+& \sum_{i=1}^4 t_{i,i+1} (d_i^\dag d_{i+1}e^{i\varphi} + d_{i+1}^\dag d_ie^{-i\varphi}) \nonumber \\
                                              &+& V (d_2^\dag d_4 + d_4^\dag d_2)
\end{eqnarray}
$H_l$ is the Hamiltonian of the leads 
\begin{equation}
H_l = \sum_{k,\alpha\in L,R} \varepsilon_{k\alpha} c^\dag_{k\alpha} c_{k\alpha},
\end{equation}
and $H_t$ is the Hamiltonian describing the tunneling between the leads and the ring
\begin{equation}
H_t = \sum_{k} (V_L c_{kL} d_1^\dag + V_R c_{kR} d_n^\dag ) + {\rm H. c.},
\end{equation}
where $\varepsilon_i$ are the on-site energy at the dots, $t_{i,i+1}$ are the nearest-neighbour hopping parameters (where $t_{45}=t_{41}$ should be understood), $V$ is the interdot hopping that couples the upper and lower arms of the interferometer, $\varphi=2\pi\Phi/4\Phi_0=\pi\Phi/2\Phi_0$ is the phase $\varphi=\pi\phi/2$ acquired due to interdot hopping in a magnetic field threading the ring with a reduced  flux $\phi=\Phi/\Phi_0$ (i.e., in units of the quantum $\Phi_0=h/e$), and $n$ is the site of contact to the right lead. The left lead is always attached to the dot 1, as shown in Fig. \ref{system}; for brevity we refer to them as connections (1,2) and (1,3).

The current through the device can be calculated with the Landauer equation
\begin{equation}
I=\frac{2e}{h}\int dE \ T(E) \left[f_L(E)-f_R(E) \right],
\end{equation}
where $f_L$ and $f_R$ are the Fermi distributions at the L and R leads. At low temperatures, the transmission function represents the dimensionless conductance (in units of the quantum $e^2/2h$) and is calculated as
\begin{equation}
T(E) = 4{\rm Tr}({\bf \Gamma}^L {\bf G}^r(E) {\bf \Gamma}^R {\bf G}^a(E)),
\end{equation}
where ${\bf G}^a$ and ${\bf G}^r$ are the matrix representation of the advanced and retarded Green functions, and ${\bf \Gamma}^L$ and ${\bf \Gamma}^R$ are the spectral densities of the leads.\\

In the {\em wide band} approximation, the Green function of the connected system is given by
\begin{equation}
G_{1n}^r = \frac{g_{1n}}{1-\Gamma^2(g_{11}g_{nn}-|g_{1n}|^2)-i\Gamma(g_{11}+g_{nn})}.
\label{connected G1n}
\end{equation}
where $g_{ij}$ is the retarded Green function of the isolated system, and $\Sigma_L=\Sigma_R=i\Gamma$ are the self-energies of the leads, considered to be energy independent.
\section{\label{section partitioning} Transmission pathways}
Partitioning of the basis space is usually employed for isolating the effects on the part of interest from the rest of the system \cite{Lowdin62}. Here, we apply a partitioning technique to recover the notion of spatial transmission pathways \cite{Hansen09}. 
The $4\times 4$ Hamiltonian can be partitioned in terms of $2 \times 2$ matrices as follows:
\begin{equation}
H = \left(\begin{array}{cc}
H^P & U \\
U^\dag & H^Q
\end{array}\right),
\label{block Hamiltonian}
\end{equation}
where $H^P=PHP$ is the part of the Hamiltonian projected on the subspace of orbitals centered on the sites of connection 1 and $n$, where $P=|1\rangle\langle 1| + |n\rangle\langle n|$, whilst $H^Q$ is the projection of $H$ on the complementary subspace $Q=1-P$. Matrices $U$ and $U^\dag$ contain matrix elements connecting states belonging to $P$ and $Q$. The Green function can be obtained by block matrix decomposition
\begin{equation}
g = (E-H)^{-1} = 
\left(\begin{array}{cc}
E-H^P & -U \\
-U^\dag & E-H^Q
\end{array}\right)^{-1}.
\end{equation}
We are interested here in the Green function projected on the subspace of the connection sites, i.e., its $P$-block. Hence, $g^P$ can be obtained from the inverse of the Schur complement of the $Q$-block, $E-H^Q$,
\begin{equation}
g^P=(E-H^P -U (E-H^Q)^{-1} U^\dag)^{-1} = (E-H^P -U g^Q U^\dag)^{-1},
\end{equation}
from which an effective Hamiltonian can be defined as 
\begin{equation}
H_{\rm eff} = E-(g^P)^{-1} = H^P +U g^Q U^\dag = H^P + \Sigma,
\end{equation}
where $\Sigma=U g^Q U^\dag$ is the self-energy that contains the interactions involving the orbitals not connected to the leads.
An approach similar to the one outlined above has been used in an analytical treatment of quantum interference in a benzene ring \cite{Hansen09}. 

The Green function written in terms of the self-energy, 
\begin{equation}
g^P(E) = \frac{1}{\Delta} \left( \begin{array}{cc}
E-\varepsilon_n-\Sigma_{nn} & t_{1n}+\Sigma_{1n} \\
t_{n1}+\Sigma_{n1} & E-\varepsilon_1-\Sigma_{11}
\end{array}\right),
\label{total Green from Heff}
\end{equation}
with $\Delta= \det(E-H_{\rm eff})=(E-\varepsilon_1-\Sigma_{11})(E-\varepsilon_n-\Sigma_{nn})-|\Sigma_{1n}|^2$, contains all the matrix elements needed for the calculation of the conductance between the sites 1 and $n$. Their poles are given by the zeroes of the secular determinant $\Delta=0$, while the antiresonances comes from zeroes in $g_{1n}=(t_{1n}+\Sigma_{1n})/\Delta$. When $t_{1n}=0$, as in the (1,3) connection, the antiresonances arise from the zeroes of $\Sigma_{1n}$.
Interestingly, the self-energy $\Sigma_{ij}=\Sigma^A_{ij}+\Sigma^B_{ij}+\Sigma^C_{ij}$ becomes a sum of contributions throughout paths from above (A), from below (B) and through the interarm coupling (C). The vanishing of the transmission occurs when the contributions from those paths interfere destructively thus cancelling the element $t_{1n}+\Sigma_{1n}$ of the effective Hamiltonian.
\\ \indent
For the (1,3) connection, the contributions become
\begin{widetext}
\begin{eqnarray}
\noindent
\Sigma^{\rm A}_{11} &=& t_{12}^2 g_{22}, \hspace{1.7cm}
\Sigma^{\rm B}_{11} = t_{41}^2g_{44} , \hspace{1.8cm}
\Sigma^{\rm C}_{11} = t_{12}g_{24} t_{41} (e^{2i\varphi}+e^{-2i\varphi}), 
\label{self-energy contributions 11}\\
\Sigma^{\rm A}_{13} &=& t_{12} g_{22} t_{23}e^{2i\varphi}, \hspace{0.7cm}
\Sigma^{\rm B}_{13} = t_{41} g_{44} t_{34}e^{-2i\varphi}, \hspace{0.5cm}
\Sigma^{\rm C}_{13} = g_{24}(t_{41}t_{23}+t_{12}t_{34}), 
\label{self-energy contributions 13}\\
\Sigma^{\rm A}_{33} &=& t_{23}^2 g_{22}, \hspace{1.7cm}
\Sigma^{\rm B}_{33} = t_{34}^2 g_{44}, \hspace{1.8cm}
\Sigma^{\rm C}_{33} = t_{23}g_{24} t_{34} (e^{2i\varphi}+e^{-2i\varphi}),
\label{self-energy contributions 33}
\end{eqnarray}
\end{widetext}
where
\begin{equation}
g_{22}=(E-\varepsilon_4)/D, \hspace{0.5cm}
g_{44}=(E-\varepsilon_2)/D, \hspace{0.5cm}
g_{24}=V/D,
\end{equation}
and $D=(E-\varepsilon_2)(E-\varepsilon_4)-V^2=(E-E_a)(E-E_b)$, with $E_a$ and $E_b$ being the energies of the bonding ($E_b$) and antibonding ($E_a$) orbitals of the molecule formed between the sites 2 and 4 described by $H^Q$, namely, $E_{a,b}=(\varepsilon_2+\varepsilon_4)/2 \pm\sqrt{V^2+(\varepsilon_2^2+\varepsilon_4^2-2 \varepsilon_2 \varepsilon_4)/4}$. 
\\ \indent
On the other hand, the conditions for the onset of peaks of resonance in the linear conductance of the ring connected to the leads, can also be analyzed from the poles of the Green functions of the disconnected device. We state that  the poles of $g_{1n}$ will show finite (or even perfect) transmission, while those poles of $g_{11}$ or $g_{nn}$ (which are not poles of $g_{1n}$) will show transmission zeros (antiresonances).

Firstly, consider a non degenerate energy eigenvalue $E_k$ of the disconnected device with eigenfunction $|\psi_k\rangle$ which is a linear combination of the site orbitals $|i\rangle$, $|\psi_k\rangle=\sum_i c_{ki} |i\rangle$, with $c_{ki}=\langle i| \psi_k \rangle$. If the state $|\psi_k\rangle$ has a non vanishing weight at the connection sites 1 and $n$ simultaneously, i.e. $c_{k1}\ne 0 \ne c_{kn}$, then the spectral representation
\begin{equation}
g_{1n}(E) = \sum_k \frac{\langle 1|\psi_k\rangle \langle\psi_k| n\rangle}{E-E_k},
          = \sum_k \frac{c_{k1} c_{kn}^{*}}{E-E_k}
\label{spectral representation}
\end{equation}
shows that $E_k$ is a pole of $g_{1n}$ because the term $c_{k1} c_{kn}^{*}/(E-E_k)$
is present in the expansion (\ref{spectral representation}). Furthermore also  the terms
\begin{equation}
\frac{c_{k1} c_{k1}^{*}}{E-E_k}, \hspace{0.5cm} {\rm and} \hspace{0.5cm}
\frac{c_{kn} c_{kn}^{*}}{E-E_k}
\end{equation}
will be present in the spectral representation of $g_{11}$ and $g_{nn}$, respectively, and $E_k$ will also be a pole of them. That is, the poles of $g_{1n}$ also become poles of $g_{11}$ and $g_{nn}$ 
and all three Green functions $g_{ij}$ diverges as  $g_{ij}(E)\approx R_{ij}/(E-E_k)$, where $R_{ij}=c_{ki} c_{kn}^{*}$ ($i,j=1,n$) is the residue of $g_{ij}$ at the simple pole $E_k$. Therefore, the Green function of the connected ring, Eq.(\ref{connected G1n}), can be approximated as 
\begin{widetext}
\begin{eqnarray}
G_{1n}(E) &\approx& \frac{R_{1n}(E-E_k)^{-1}}{1-\Gamma^2(R_{11}R_{nn}-R_{1n}^2)(E-E_k)^{-2}-i\Gamma(R_{11}+R_{nn})(E-E_k)^{-1}} \nonumber \\ 
  &=&  \frac{R_{1n}}{(E-E_k)-\Gamma^2(R_{11}R_{nn}-R_{1n}^2)(E-E_k)^{-1}-i\Gamma(R_{11}+R_{nn})}.
\label{Green near poles}
\end{eqnarray}
\end{widetext}
Taking into account that $R_{11}R_{nn}-R_{1n}^2=|\langle 1|\psi_k\rangle|^2 |\langle n|\psi_k\rangle|^2 - |\langle 1|\psi_k\rangle \langle \psi_k|n \rangle|^2 = 0$, it reduces to
\begin{equation}
G_{1n} \approx \frac{R_{1n}}{(E-E_k)-i\Gamma(R_{11}+R_{nn})} \stackrel{E\rightarrow E_k}{\longrightarrow} \frac{iR_{1n}}{\Gamma(R_{11}+R_{nn})}
\end{equation}
which shows that the transmission has a pole at $E=E_k+i\Gamma(R_{11}+R_{nn})$ and a finite transmission $T_{1n}= 4R^2_{1n}/(R_{11}+R_{nn})^2$. The pole $E_k$ acquires a finite width proportional to the coupling to the leads $\Gamma$. In the particular case where the sites 1 and $n$ are topologically equivalent because of the symmetry of the system, $R_{11}=R_{nn}=R_{1n}$ so that perfect transmission occurs.

On the other hand, if $E=E_k$ is a pole of $g_{11}$ or $g_{nn}$, but not of $g_{1n}$, the numerator $g_{1n}(E_k)$ of Eq. (\ref{connected G1n}) is finite whilst its denominator diverges; therefore $T_{1n}$ will show an antiresonance at $E=E_k$.

In other words, a finite transmission occurs when the eigenstate $\psi_k$ of the isolated system have non-vanishing projection on the orbitals $|1\rangle$ and $|n\rangle$: $\langle 1|\psi_k\rangle$ and $\langle n|\psi_k\rangle$. Reciprocally, if one of them equals zero, the electron of energy $E=E_k$ has a vanishing probability of being at both sites, and therefore no transmission can occur.
\\ \indent
The case when the eigenvalue $E_k$ is degenerate requires a modification of the above argument. In such a case, there are more than one states $\psi_k^{(1)}, \psi_k^{(2)}\ldots \psi_k^{(p)}$ having the same energy $E_k$. For the sake of simplicity, consider just two degenerate eigenfunctions $|\psi_k^{(p)}\rangle=\sum_i c_{ki}^{(p)} |i\rangle$, ($p=1,2$). The spectral representation of the Green function now reads
\begin{equation}
g_{ij}(E) = \sum_k \frac{R_{ij} }{E-E_k}, \hspace{0.5cm}
R_{ij} = c_{ik}^{(1)} c_{jk}^{(1)*} + c_{ik}^{(2)} c_{jk}^{(2)*}.
\label{spectral representation degenerate}
\end{equation}
The property $R_{11}R_{nn}-R_{1n}^2= 0$, valid for the nondegenerate case, does no longer hold here. Instead
\begin{eqnarray}
R_{11}R_{nn}-R_{1n}^2 &=& (c_{1k}^{(1)})^2 (c_{nk}^{(2)})^2 + (c_{1k}^{(2)})^2 (c_{nk}^{(1)})^2 \nonumber \\
                      &&- c_{1k}^{(1)} c_{nk}^{(1)} c_{1k}^{(2)*} c_{nk}^{(2)*} - c_{1k}^{(2)} c_{nk}^{(2)} c_{1k}^{(1)*} c_{nk}^{(1)*},\nonumber \\
\label{condition degenerate}
\end{eqnarray}
which can vanish or not depending on the $c_{1k}^{(p)}$ and $c_{nk}^{(p)}$. If $R_{11}R_{nn}-R_{1n}^2 = 0$, all above discussion holds; if not, the real part of the denominator in Eq. (\ref{Green near poles}) diverges near $E=E_k$ as $\Gamma(R_{11}R_{nn}-R_{1n}^2)/(E-E_k)$ and the transmission becomes suppressed ($T=0$).
\section{\label{section results} Results and discussion}
\begin{figure}[!ht]
\includegraphics[width=8cm]{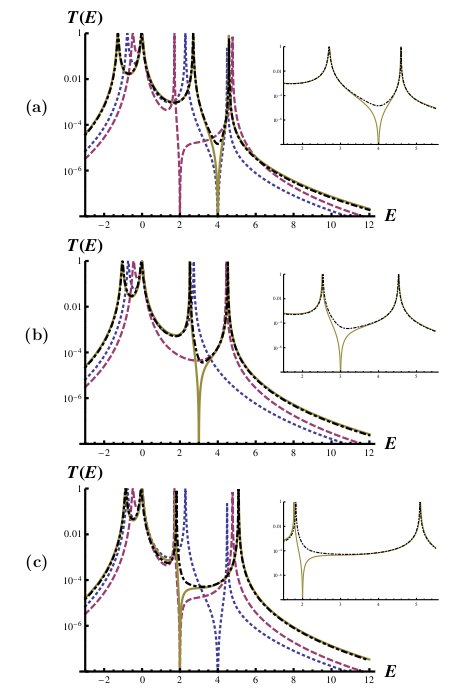}
\caption{\label{T(E) symmetric ring} Transmission function of a ring with interarm coupling in the connection (1,3) (solid line), and three-sites chains with a laterally coupled side dot representing the upper (dashed line) and lower (dotted line) arms as a function of the energy $E$. All dots are connected by the same interdot hopping $t_{12}=t_{23}=t_{34}=t_{41}=1$ and interarm couplings $V=-1, 0, 1$, without magnetic flux (solid line) and with a flux $\Phi=0.1\Phi_0$ (dot-dashed line). (a) Ring with coupling $V=-1$, (b) ring with disconnected arms ($V=0$), and (c) ring with coupling $V=1$. The $V$ couplings were chosen to show the tuning of the antiresonance with the one for the single chain with a lateral dot. The insets shows, in more detail, the suppression of the antiresonance due to the applied magnetic flux.}
\end{figure}

Firstly, consider a ring threaded by a magnetic flux with all interdot hopping parameters equal to each other ($t_{12}=t_{23}=t_{34}=t_{41}=t$) and interarm coupling $V$. The on-site energies are taken as $\varepsilon_1=\varepsilon_3=0$, $\varepsilon_2=2$ and $\varepsilon_4=4$. All the results are obtained with a coupling to the leads $\Gamma=0.05$. Figure \ref{T(E) symmetric ring} shows the dimensionless conductance $T(E)$ for the ring connected to the leads in the configuration (1,3) (solid line), and for the three-sites chains forming the upper (dashed line) and lower (dotted line) arms with the side-dot 4 and 2, respectively, coupled by $V$. The transmission with an applied magnetic flux $\Phi=0.1\Phi_0$ is shown in dot-dashed line. 
The transmission show four peaks at the energy eigenvalues of the systems. The ring with disconnected arms Figure \ref{T(E) symmetric ring}(b) shows also an antiresonance, not present in the chains because for $V=0$ there is no side-coupled dot. This suppression of the transmission in the ring is due to the cancellation of the contributions to the self-energy throughout the upper and lower paths ($\Sigma_{13}=\Sigma_{13}^A+\Sigma_{13}^B=0$). 
In general, eqs. (\ref{self-energy contributions 13}) show that the self-energy vanishes if 
\begin{eqnarray}
&&\Sigma_{13}=t^2(g_{22} e^{2i\varphi}+g_{44} e^{-2i\varphi}+2g_{24}) \nonumber \\
&&           =\frac{t^2}{D}\left[ (E-\varepsilon_4)e^{2i\varphi}+(E-\varepsilon_2)e^{-2i\varphi}+2V\right]=0
\end{eqnarray}
In absence of magnetic flux ($\varphi=0$), $\Sigma_{13}$ vanishes at the energy $E=\bar{\varepsilon}-V=(\varepsilon_2+\varepsilon_4)/2-V$.
Figures \ref{T(E) symmetric ring}(a)-\ref{T(E) symmetric ring}(c) depicts the tuning of the antiresonance with $V$, for $V=(\varepsilon_2-\varepsilon_4)/2$, $V=0$ and $V=(\varepsilon_4-\varepsilon_2)/2$, respectively, such that the antiresonance of the ring can be made to coincide with that from the upper and lower arms with a lateral dot, respectively.
When there is a finite magnetic flux, $\Sigma_{13}$ is complex and its cancellation requires vanishing its real and imaginary parts, i.e, $(2E-\varepsilon_2-\varepsilon_4)\cos 2\varphi+2V=0$, and $(\varepsilon_2-\varepsilon_4)\sin 2\varphi=0$. Both equations cannot be satisfied simultaneously, except when $\varepsilon_2=\varepsilon_4$, which presents an antiresonance at $\varepsilon_2-V/\cos 2\varphi$. Therefore, the magnetic field eliminates the antiresonance for a ring with different site energies $\varepsilon_2 \ne \varepsilon_4$ for arbitrary $V$. This suppression of the antiresonance is shown in the insets of Figures \ref{T(E) symmetric ring}(a)-\ref{T(E) symmetric ring}(c) for a flux $\Phi=0.1\Phi_0$.

\begin{figure}[!ht]
\includegraphics[scale=0.5]{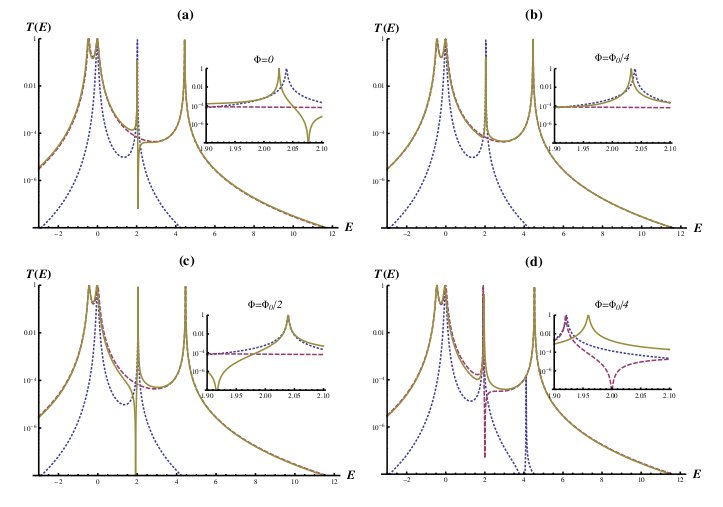} 
\caption{\label{T(E) Fano} Transmission for the connection (1,3) as a function of the Fermi energy for an asymmetric ring with hopping parameters $t_{12}=t_{23}=0.2$ (upper arm), $t_{34}=t_{41}=1$ (lower arm) and on-site energies $\varepsilon_2=2$ and $\varepsilon_2=4$ (solid line), as compared to the transmissions through the upper arm only (dotted blue line), and throughout the lower arm only (dashed red line). Figures (a)-(c) correspond to a decoupled ring ($V=0$) (a) at zero magnetic field, (b) at $\Phi=\Phi_0/4$, and (c) magnetic flux $\Phi=\Phi_0/2$. Figure (d) has an interarm coupling $V=0.5$ and magnetic flux $\Phi=\Phi_0/2$. The insets show in more detail the behaviour of the curves around the resonance.}
\end{figure}

Application of gate potentials at dots 2 and 4 allows to tune their on-site energies. Figure \ref{T12(E,E_dot) symmetric hopping} shows the effect on the transmission of varying these parameters and the electron energy $E$ with and without magnetic flux for the connection (1,2). Bright lines and dark regions represent zones of high and low transmission, respectively. Variation of $\varepsilon_2$ (left panels) was done at $\varepsilon_4=4$ while variation of $\varepsilon_4$ (right panels) was done at $\varepsilon_2=2$. In absence of magnetic field (upper panels) three peaks of conductance are visible depicted by the bright curves. The central peak has a linear dependence on the energies $\varepsilon_2$ and $\varepsilon_4$ while the external ones are weakly dependent on them, as seen from the slope of the curves. Two antiresonances are also visible, namely, a faint vertical dark line at $E=0$ independent on $\varepsilon_i$, and a vertical dark thicker line at $E=\varepsilon_4-V$ (upper left panel) and the straight line $E=\varepsilon_4-V$ (upper right panel). Hence, both $\varepsilon_2$ and $\varepsilon_4$ are equally suitable for tuning the maxima of transmission, but only the site 4, which is not connected to the leads is efficient for tuning the antiresonances. The lower panels of Figure \ref{T12(E,E_dot) symmetric hopping} show the effect of switching on a magnetic flux. As discussed for the connection (1,3), the antiresonances are cancelled out. In particular, it should be noted that the antiresonance at $E=0$ turned into a peak of transmission. Also the second antiresonance at $E=\varepsilon_4-V$ becomes weakened and the dark sharp straight line turns into a diffuse dark region of low transmission.
\begin{figure}
\includegraphics[scale=0.5]{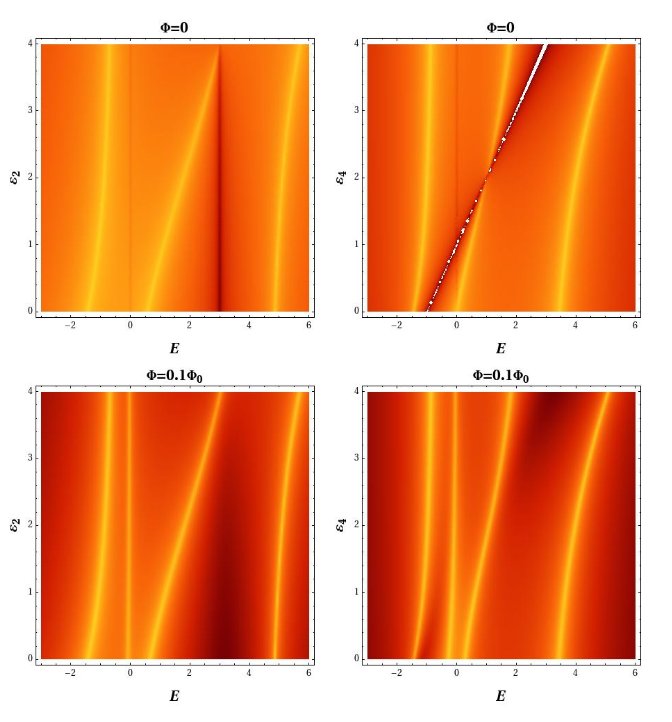}
\caption{\label{T12(E,E_dot) symmetric hopping} The logarithm of the transmission function of a ring with coupling between the arms in the connection (1,2) as a function of the energy $E$ and the dots energies $\varepsilon_2$ and $\varepsilon_4$ with and without magnetic flux. All dots are connected by the same interdot hopping parameters $t_{12}=t_{23}=t_{34}=t_{41}=V=1$, and the magnetic flux was taken as $\Phi=0.1\Phi_0$. Bright lines and dark regions represent zones of high and low transmission, respectively. Variation of $\varepsilon_2$ (left panels) was done at $\varepsilon_4=4$ while variation of $\varepsilon_4$ (right panels) was done at $\varepsilon_2=2$.} 
\end{figure}
Figure \ref{T(E) Fano} shows in solid lines the transmission through an asymmetric ring where the upper arm has hoppings much smaller than those of the lower arm ($t_{12}=t_{23}\ll t_{34}=t_{41}$). The dotted blue and dashed red lines are the transmissions of the upper and lower arms separately, calculated as three-site chains with a central site having energy $\varepsilon_2$ and $\varepsilon_4$, respectively, and lateral sites having on site energies $\varepsilon=0$. 
In Figure \ref{T(E) Fano}, the four peaks of conductance, corresponding to connection (1,3), can be recognized as those from the lower arm along with the on-site energy of site 2. The transmission of the ring nearly coincides in almost the whole range with that of the lower arm, thus showing that the conduction is throughout such a pathway at almost every energy, except for $E\approx \varepsilon_2$. When the incident electron is resonant with the site 2, the transmission is well described by the resonant peak of the upper pathway. Nevertheless, none of the paths by themselves can provide the onset of the antiresonance close to $E\approx \varepsilon_2$. 
The self energy $\Sigma_{13}=\Sigma_{13}^{\rm A}+\Sigma_{13}^{\rm B}\approx \Sigma_{13}^{\rm B}$ because $\Sigma_{13}^{\rm A}=t_{\rm up}^2/(E-\varepsilon_2)\ll\Sigma_{13}^{\rm B}= t_{\rm down}^2/(E-\varepsilon_4)$, except for $E\approx\varepsilon_2$ when they can become comparable. Hence, in a neighbourhood of $E\approx\varepsilon_2$, the self energy can be approximated as $\Sigma_{13}=t_{\rm up}^2/(E-\varepsilon_2)+t_{\rm down}^2/(\varepsilon_2-\varepsilon_4)$ which vanishes for $E=\varepsilon_2+t_{\rm up}^2 (\varepsilon_4-\varepsilon_2)/t_{\rm down}^2$, that is, slightly to the right of the peak $E=\varepsilon_2$, thus giving the Fano-like profile.
\\ \indent
The Fano-like peak shows the signature of the interference between both paths, typical when a localized state interferes with a continuum. Figures \ref{T(E) Fano}(a)-\ref{T(E) Fano}(c) show the dependence of the Fano profile with the magnetic flux. With no magnetic flux, there is the above discussed Fano resonance due to the path interference; the application of a flux $0<\Phi<\Phi_0/4$ suppress the antiresonance leaving only a dip in the transmission which also disappears at $\Phi=\Phi_0/4$ leaving only the resonant peak, as seen in \ref{T(E) Fano}(b). Further increase of the flux in the range $\Phi_0/4 <\Phi< \Phi_0/2$ produce a new dip at an energy slightly smaller than $\varepsilon_2$ while moves the peak to energies slightly higher. At $\Phi=\Phi_0/2$ the dip in the transmission of the ring becomes an antiresonance, with the resonant peak tuned with that of the upper chain as seen in the inset of figure \ref{T(E) Fano}(c). Between $\Phi_0/2$ and $\Phi_0$, the behaviour of $T(E)$ is reversed, such that a cycle is completed in a period of $\Phi_0$. Figure \ref{T(E) Fano}(d) depicts the transmission through the ring with an interarm coupling $V=0.5$. Now the transmission through the lower (dashed line) and upper (dotted line) arms, including the site laterally coupled by $V$, shows Fano-like resonances at $\varepsilon=\varepsilon_2$ and at $\varepsilon=\varepsilon_4$, respectively. The transmission through the ring (solid line) still remains close to that of the lower arm with a lateral connection to site 2. Then, the overall picture for the transmission through a ring with different connection strengths along each arm, is that of the transmission throughout the stronger pathway (i.e., the one with larger hoppings) at almost every energy, except at the one resonant with the energy of the site connecting the arms where a Fano interference occurs.

In the presence of a magnetic flux, the self energy is a complex quantity and its modulus should be considered. Eqs. (\ref{self-energy contributions 13}) shows that the flux $\Phi$ introduces a phase $\pm2\varphi$ in the paths throughout the arms while no phase change occurs in the self-energy corresponding to the interarm coupling. Let us call $\Sigma_{13}(\varphi)$ the self energy with magnetic field. Then, $\Sigma_{13}(\varphi)=\Sigma_{13}^{\rm A}e^{2i\varphi}+\Sigma_{13}^{\rm B}e^{-2i\varphi}+\Sigma_{13}^{\rm C}$, where $\Sigma_{13}^{\rm A,B,C}$ are the real self-energies at zero magnetic field, such that
\begin{eqnarray}
|\Sigma_{13}(\varphi)|^2 &=& (\Sigma_{13}^{\rm A})^2 + (\Sigma_{13}^{\rm B})^2 + (\Sigma_{13}^{\rm C})^2 + 2 \Sigma_{13}^{\rm A} \Sigma_{13}^{\rm B} \cos 4\varphi \nonumber \\ 
&&+ 2 (\Sigma_{13}^{\rm A}+\Sigma_{13}^{\rm B}) \Sigma_{13}^{\rm C}\cos 2\varphi,
\end{eqnarray}
where the first three terms represent the non-interfering transmission along the paths A, B and C. The last two terms contain the effect of the interference due to the quantum and magnetic phases. It is clearly noted that even for $\varphi=0$ there is an interference between the path contributions to the self-energy. Interestingly, there are two periods in the magnetic phase; a period $\Phi=\Phi_0$ (associated to $\cos 4\varphi$) and a period $\Phi=2\Phi_0$ (associated to $\cos 2\varphi$). When there is no interarm coupling, $\Sigma_{13}^{\rm C}=0$, the latter is not present. On the other hand, as soon as a finite $V$ exists, the self-energy acquires the longer period modulated by the shorter one. Such a behaviour has been observed in experiments \cite{Ihn07} and were termed as Fano resonances of the {\em big} and {\em small orbits}. 
Figure \ref{T(E,phi)} shows the transmission (in a log scale) $T(E,\varphi)$ as a function of the energy and the magnetic flux, for various values of the interarm coupling $V=0,$ 0.5 and 1, and for the two ways of connecting to the leads. 
The top (left and right) panels, corresponding to $V=0$, are the only ones showing a period $\Phi_0$ in the flux. As $V$ increases, the period $2\Phi_0$ becomes apparent. Finally, the bottom (left and right) panels show the transmission for a single subring obtained from decoupling the sites 2 and 3 (i.e, $t_{23}=0$), as also done in the experiments \cite{Ihn07}, where the period of the small orbit is clearly apparent.
\begin{figure}[!ht]
\includegraphics[scale=0.5]{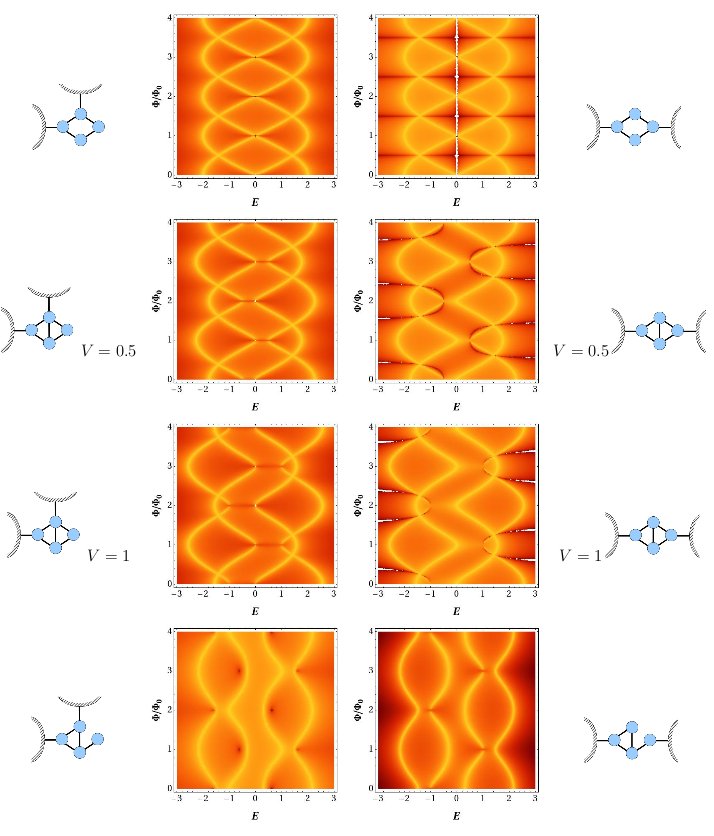} 
\caption{\label{T(E,phi)} The logarithm of the transmission coefficient $\log(T)$ as a function of the electron energy $E$ (horizontal axis) and the magnetic flux $\Phi$ (vertical axis) given in units of the quantum of flux $\Phi_0$, for $V=0$, 0.5, and 1. All hopping parameters are $t_{i,i+1}=1$ and all site energies $\varepsilon=0$. The case $V=0$ corresponds to that of a single loop ({\em big orbit}) enclosing the magnetic flux. The bottom pictures correspond to a single subring with $V=t_{12}=t_{34}=t_{41}=1$ and $t_{23}=0$ ({\em small orbit})}
\end{figure}
In the configuration (1,2) there are three interfering paths, namely, the direct path through sites 1 and 2 ($t_{12}$), the path $\Sigma_{12}^{{\rm C}}$ ($1\rightarrow4\rightarrow2$) and the path $\Sigma_{12}^{{\rm B}}$ ($1\rightarrow4\rightarrow3\rightarrow2$), contributing to the self-energy $\Sigma_{12}(\varphi)=t_{12}e^{i\varphi}+Vt_{14} g_{44} e^{-i\varphi}+t_{14} g_{43} t_{31} e^{-3i\varphi}$. 
The absolute square of the self-energy turns out
\begin{eqnarray}
\Sigma_{12}&=& t_{12}^2 + (\Sigma_{12}^{{\rm B}})^2 + (\Sigma_{12}^{{\rm C}})^2 + 2 (t_{12}+\Sigma_{12}^{{\rm B}}) \Sigma_{12}^{{\rm C}} \cos 2\varphi \nonumber \\ 
                      &&+ 2 t_{12} \Sigma_{12}^{{\rm B}} \cos 4\varphi,
\end{eqnarray}
where the phase of the big orbit ($4\varphi$) characterize the interference between the pathways along the arms of the ring, whilst the phase of the small orbit ($2\varphi$) corresponds to the interference between them and the molecular bond. The bottom panels of Figure \ref{T(E,phi)}, shows the conductance when the hopping $t_{23}$ has been set equal to zero, such that there is a single small orbit. Nevertheless, it should be noted a few differences visible as dark spots; they correspond to antiresonances determined by the different topology of the connections.

\section{\label{section conclusions} Conclusions}
We have shown the conditions for the onset of resonances and antiresonances in an artificial quantum dot molecule embedded in the arms of an Aharonov-Bohm interferometer, and its dependence on the tunable parameters.
In general, the peaks of conductance are located at the energy eigenvalue of the isolated device. A partitioning technique enables to decompose the transmission as a sum along interfering pathways. The tunability of the coupling between the arms of the interferometer, allows one to weaken or enhance the contribution through the bond of the artificial molecule. The experimentally observed change in the period of the Aharonov-Bohm phase from one to twice the quantum of flux is interpreted as due to the opening of one transmission pathway. Our results suggest that a modified coherent single-mode picture including the electron reflection in the subrings forming the small orbits, could also be of help in interpreting the experiments.
Application of a magnetic flux leads to a suppression of the antiresonances due to the partial cancellation of the destructive quantum interference. We have also discussed the differences with a connection not realized experimentally in which one of the dots of the molecule is attached to one lead. Such a configuration could provide other alternatives for tuning the conductance.

\section*{Acknowledgements} This work was partly supported by SGCyT (Universidad Nacional del Nordeste), National Agency ANPCYT and CONICET (Argentina) under grants PI 112/07, PICTO-UNNE 204/07 and PIP 11220090100654/2010. 


\end{document}